\pdfoutput=1

\documentclass[
    ,final            
  ]
  {aipproc}

\layoutstyle{6x9}

\newcommand{\oralquest}[1]{\iffalse{#1}\fi}

\graphicspath{{figures/}}

\begin{document}

\title{NEUT Pion FSI}


\author{Patrick de Perio}{
  address={University of Toronto, Department of Physics, Toronto, Ontario, Canada}
}

\begin{abstract}
The pion final state interaction model in NEUT is described. 
Modifications and tuning of the model are validated against neutrino and 
non-neutrino (pion scattering and photoproduction) data. A method for evaluating the uncertainties
in the model and propagation of systematic errors in a neutrino oscillation experiment is described, 
using T2K as a specific example.
\end{abstract}

\maketitle

\section{Introduction}

Neutrino interaction generators play an integral role in  
neutrino-nucleus scattering and oscillation experiments. The vertex of a quasi-free 
interaction on a bound nucleon, assuming an impulse approximation, is referred to as the 
primary vertex. The final state hadrons produced at this vertex must propagate through the 
nucleus before observation in a detector. These particles can interact 
via the strong force, so there is a significant probability of re-interactions, or 
final state interactions (FSI), which affect the observable final state via 
absorption, scattering and particle production. This can 
obscure the true interaction mode of the primary vertex. 
Thus, recent experiments categorize measurements by the final state, instead of assuming some 
model to extract information about the primary vertex. This necessarily convolutes the physics 
of the primary vertex and FSI, allowing model builders to fit their own neutrino interaction and 
FSI models.


Our understanding of FSI feeds 
directly into neutrino oscillation experiments such as T2K~\cite{bib:t2k_experiment}. The 
incident neutrino energy is required for precisely measuring oscillation parameters. 
A charged-current quasi-elastic (CCQE) measurement is useful
since one can assume simple 2-body kinematics 
for reconstructing the neutrino energy from the outgoing lepton. 
However, CC1$\pi$ interactions with $\pi$ absorption, 
can contaminate a CCQE measurement if the outgoing 
nucleons are undetectable, as is typically the case in \v{C}erenkov detectors. The 
resulting energy spectrum would be distorted due to 
missing energy. Also, $\pi^0$ production in neutral current (NC) interactions 
can mimic the electro-magnetic signature of an electron if one of the decay 
photons is not observed. This would contaminate a $\nu_e$ CCQE measurement for 
a $\nu_e$ appearance analysis. This motivates the need for an accurate FSI model
that also has a good handle on systematic uncertainties.

This note describes recent modifications to the $\pi$ FSI model in the 
NEUT generator~\cite{bib:neut_hayato} that improve the agreement with $\pi$ 
scattering data. The data was then used for constraining model parameter uncertainties. 
A new reweighting method, which facilitates the propagation of
these uncertainties into an oscillation analysis, is described. The new model and reweighting 
has been tested against SK atmospheric neutrino data and used in the T2K $\nu_{e}$ 
appearance analysis~\cite{bib:t2k_nueapp1st}. 

\section{Tuning and Validation of the Pion Cascade Model}
\label{ch:cascade}

The NEUT $\pi$ FSI model is a microscopic cascade that propagates the $\pi$ in finite
steps through the nucleus. The mean free path (MFP) of various intranuclear mechanisms 
are calculated from a Delta-hole model~\cite{salcedo:pionfsi} 
at low energy (LE), $p_{\pi} \le 500$~MeV/c, and from free $\pi{p}$ scattering 
cross sections at high energy (HE), $p_{\pi} > $~500~MeV/c. The MFPs are density (position) 
dependent and a Woods-Saxon distribution is assumed with parameters measured from electron 
scattering data.
The microscopic method provides a connection between $\pi$ scattering measurements, 
where the $\pi$ originates from outside the nucleus, to neutrino experiments where 
the $\pi$ is produced within the nucleus. With $\pi$ scattering data, we can tune and constrain 
the FSI model, without the complication of any additional primary vertex physics.

\begin{figure}
     
  $\begin{array}{cc}
    \resizebox{3.1in}{!}{\includegraphics{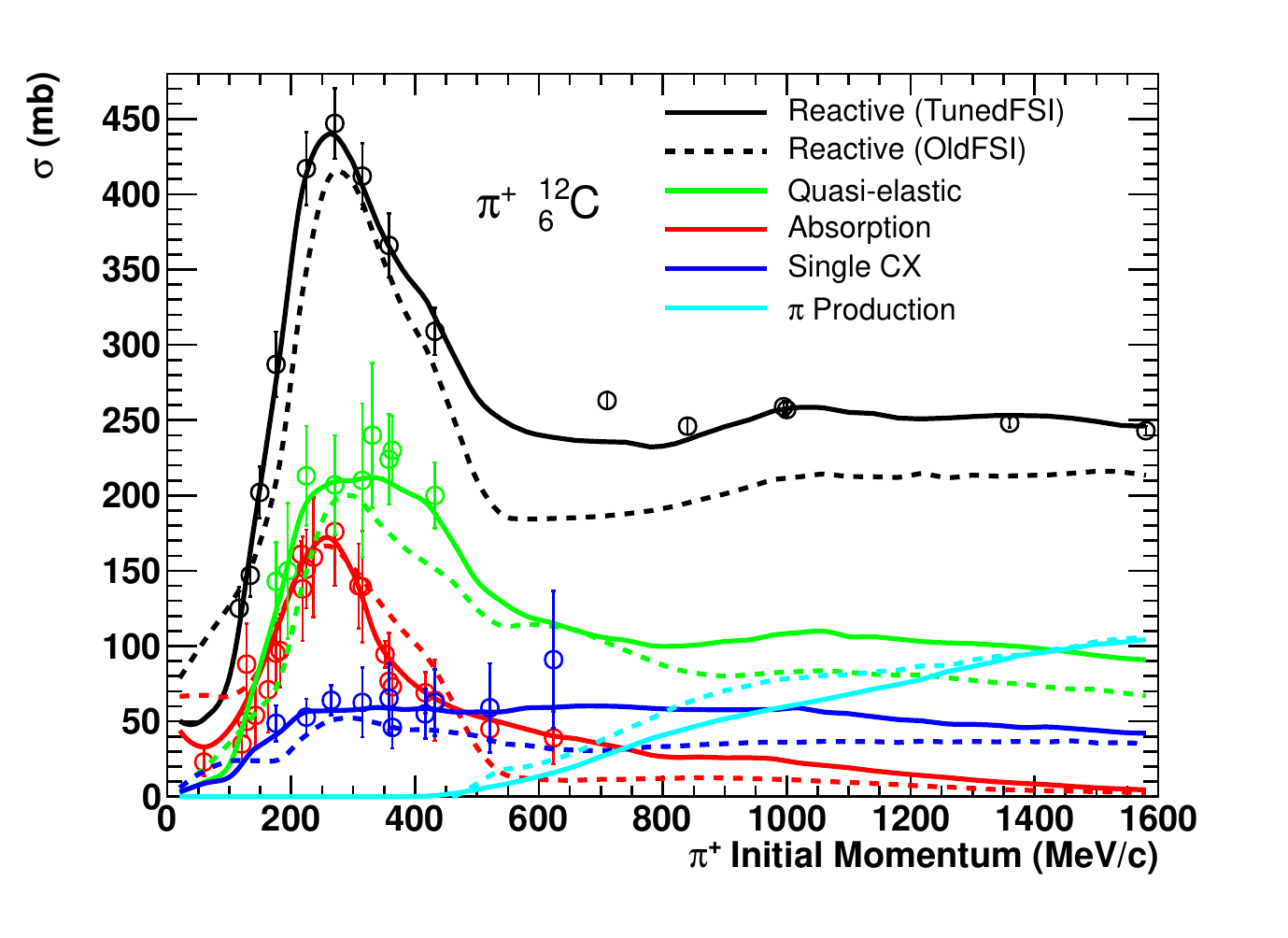}} &
    \resizebox{3.1in}{!}{\includegraphics{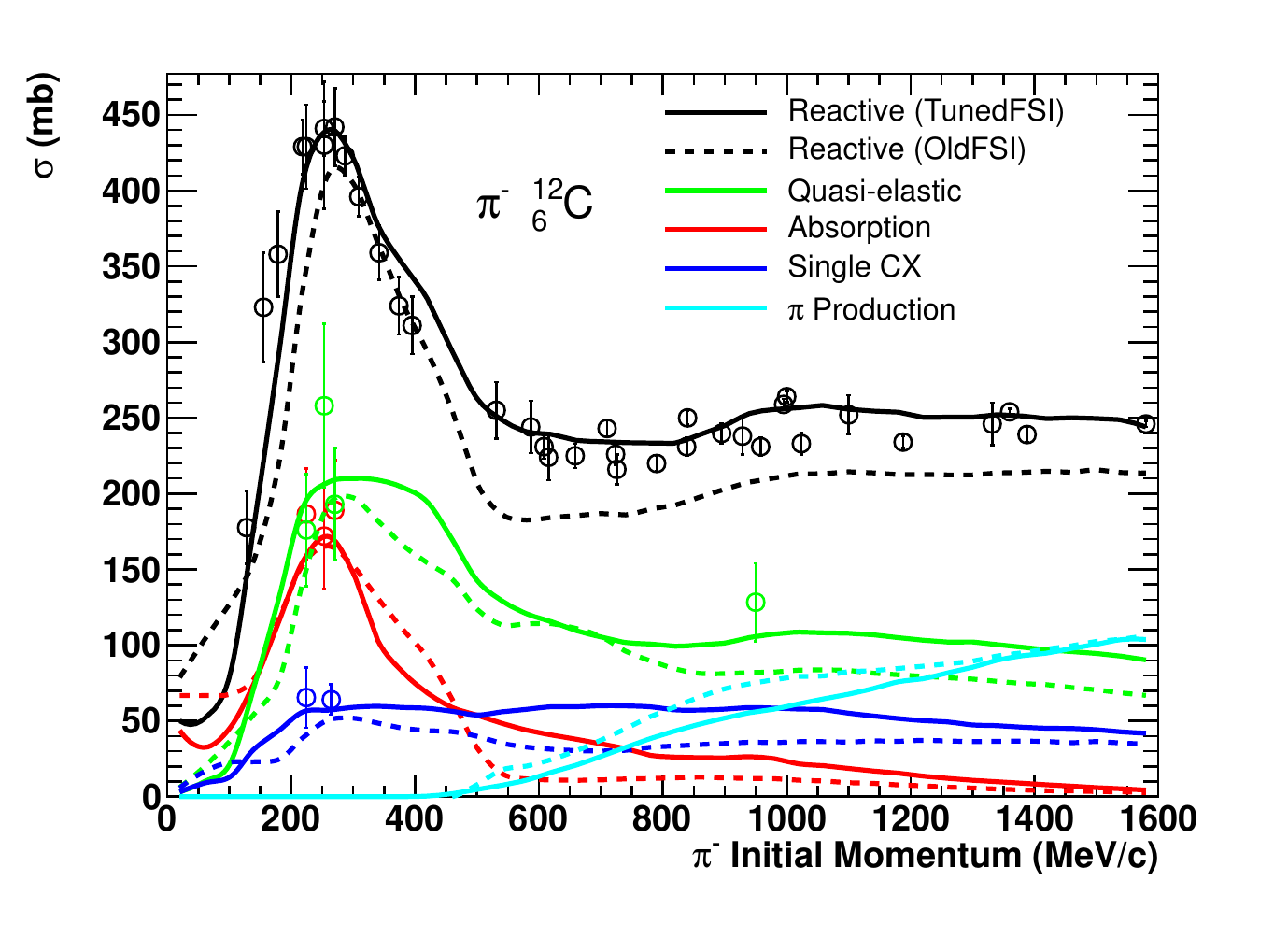}} \\
  \end{array}$  
 \vspace{-1in}
  \caption
{NEUT simulation of $\pi^{12}$C scattering compared to data 
from~\cite[e.g.][]{fujii:piscat, jones:piscat, ashery:piscat, allardyce:piscat, lee:pisummary} }
  \label{fig:piscat_c}
\end{figure}

At LE, the microscopic MFPs for absorption and scattering, were tuned from 
the original calculation so that the resulting cross sections agreed well with $\pi ^{12}$C 
scattering data in Fig.~\ref{fig:piscat_c}. A ``scattering'' vertex within the nucleus
can be either QE-like (same charge out) or single charge exchange (SCX), with the 
relative fraction determined by isospin. Note the simultaneous improved agreement 
in all channels.

\begin{figure}  

  $\begin{array}{cc}
    \resizebox{3in}{!}{\includegraphics{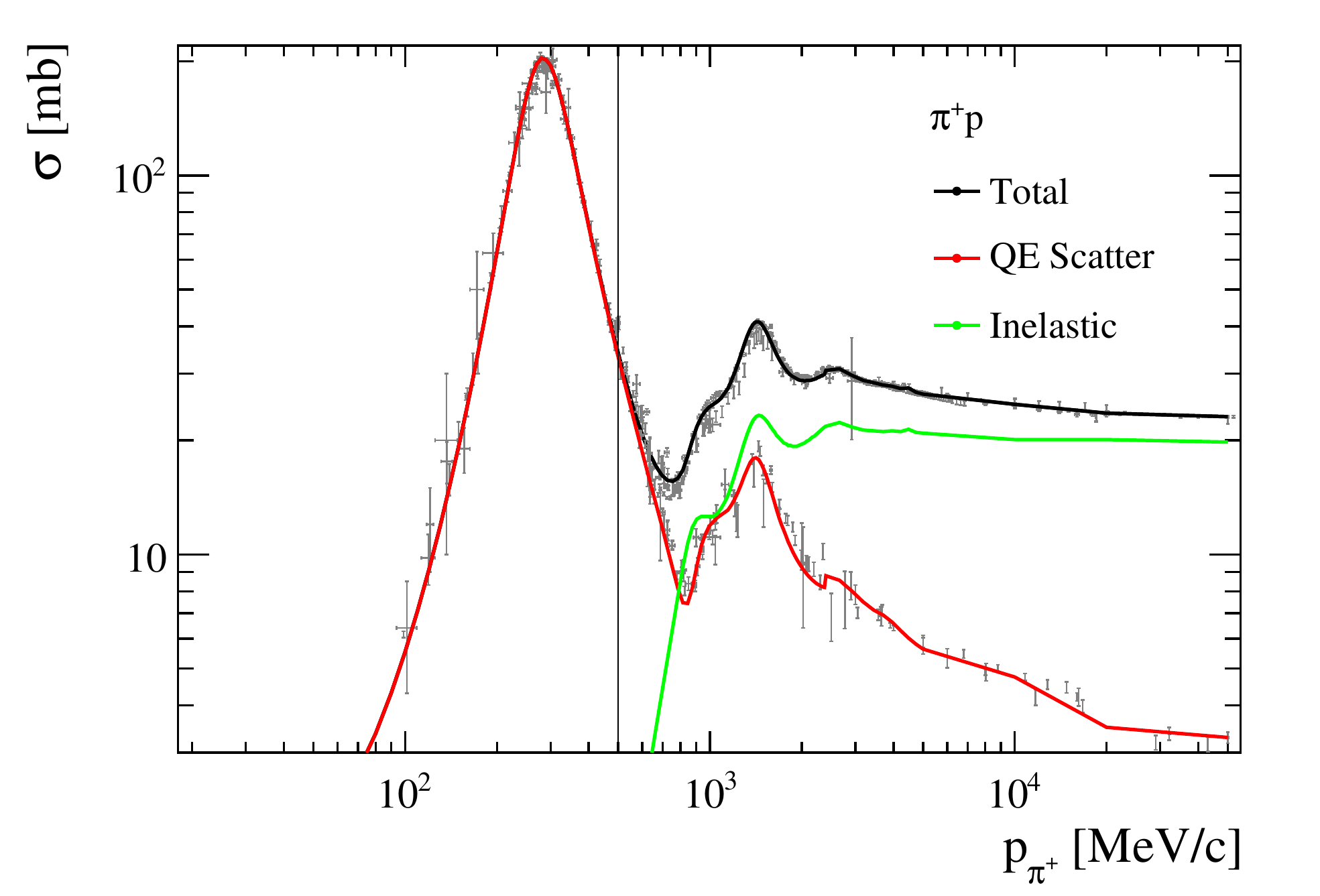}} &
    \resizebox{3in}{!}{\includegraphics{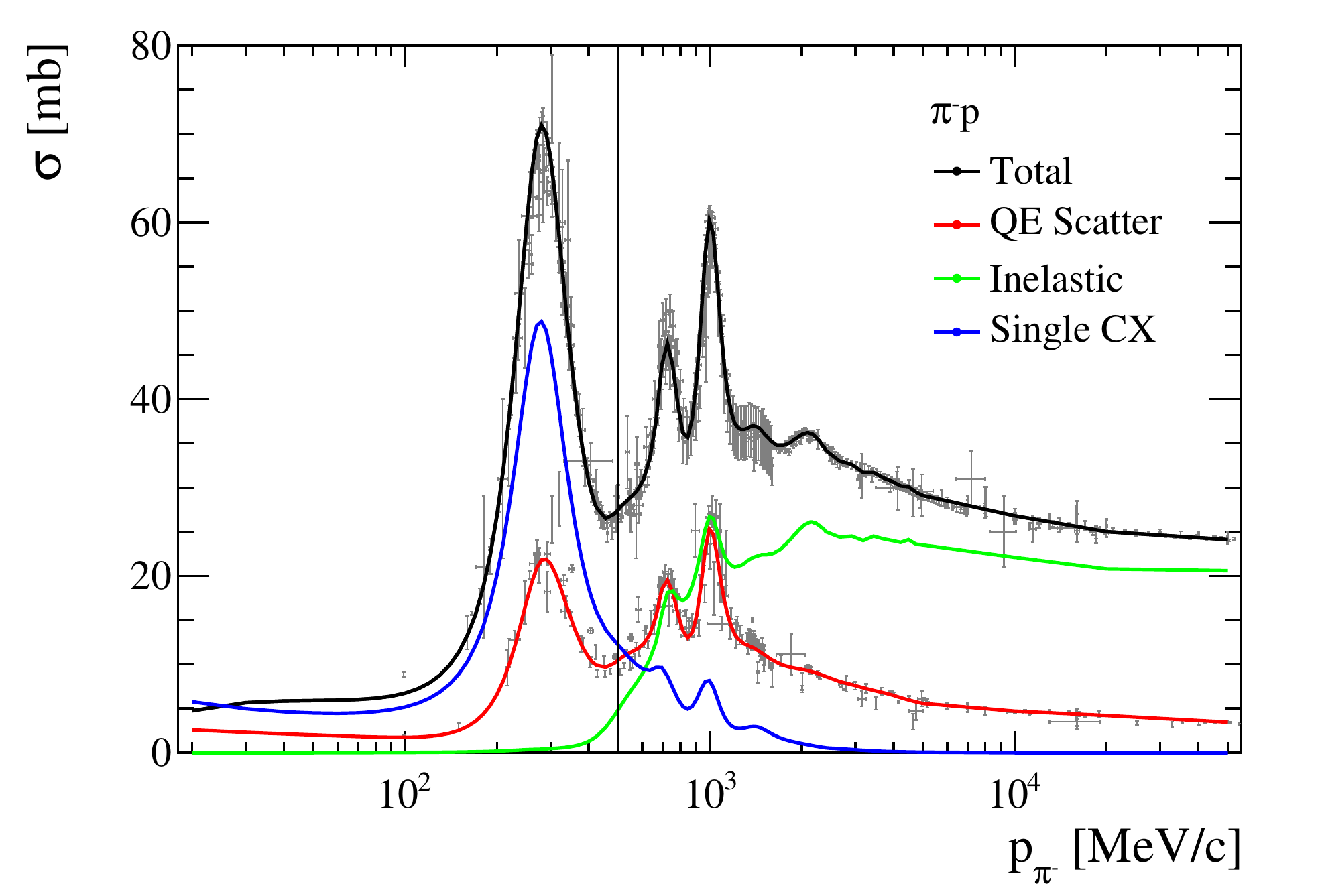}} \\
  \end{array}$  
  
  \caption
{$\pi^{\pm}$ on free proton scattering cross sections. Data from the PDG and fits by SAID~\cite{pwa:pin}.}
  \label{fig:freepxsec}
\end{figure}

At HE, the original NEUT model assumed an iso-scalar nucleus and defined the 
cross section (used for calculating the microscopic MFP) for scattering on a nucleon 
within the nucleus as: $\sigma_{QE} = \frac{1}{2}\sigma_{\pi{d}}\left(\frac{\sigma_{QE}^{free}}{\sigma_{tot}^{free}}\right)$, 
where $\sigma_{\pi{d}}$ is the total cross section for $\pi$ scattering on deuteron, 
and $\sigma_{QE}^{free}$ and $\sigma_{tot}^{free}$ are the cross sections on free proton 
as shown in Fig.~\ref{fig:freepxsec}. The $\pi^+{p}$ ($\pi^-{p}$) cross sections are also used for 
$\pi^{-}n$ ($\pi^{+}n$, $\pi^0{N}$) initial states, motivated by isospin symmetry. For 
hadron production: 
$\sigma_{had} = \frac{1}{2}\sigma_{\pi{d}}\left(1-\frac{\sigma_{QE}^{free}}{\sigma_{tot}^{free}}\right)$.

\begin{figure}
     
  $\begin{array}{cc}
    \resizebox{3in}{!}{\includegraphics{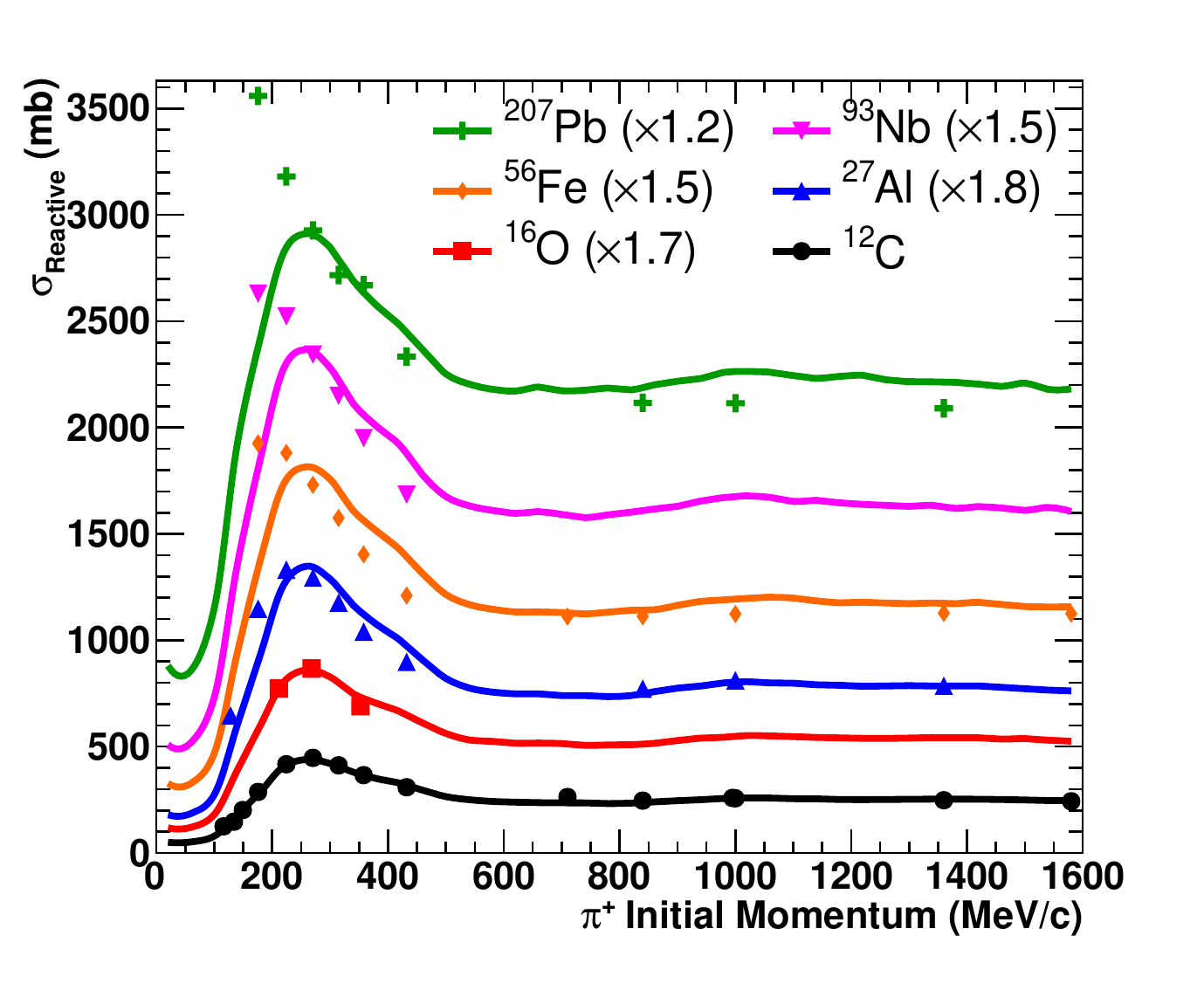}} &
    \resizebox{3in}{!}{\includegraphics{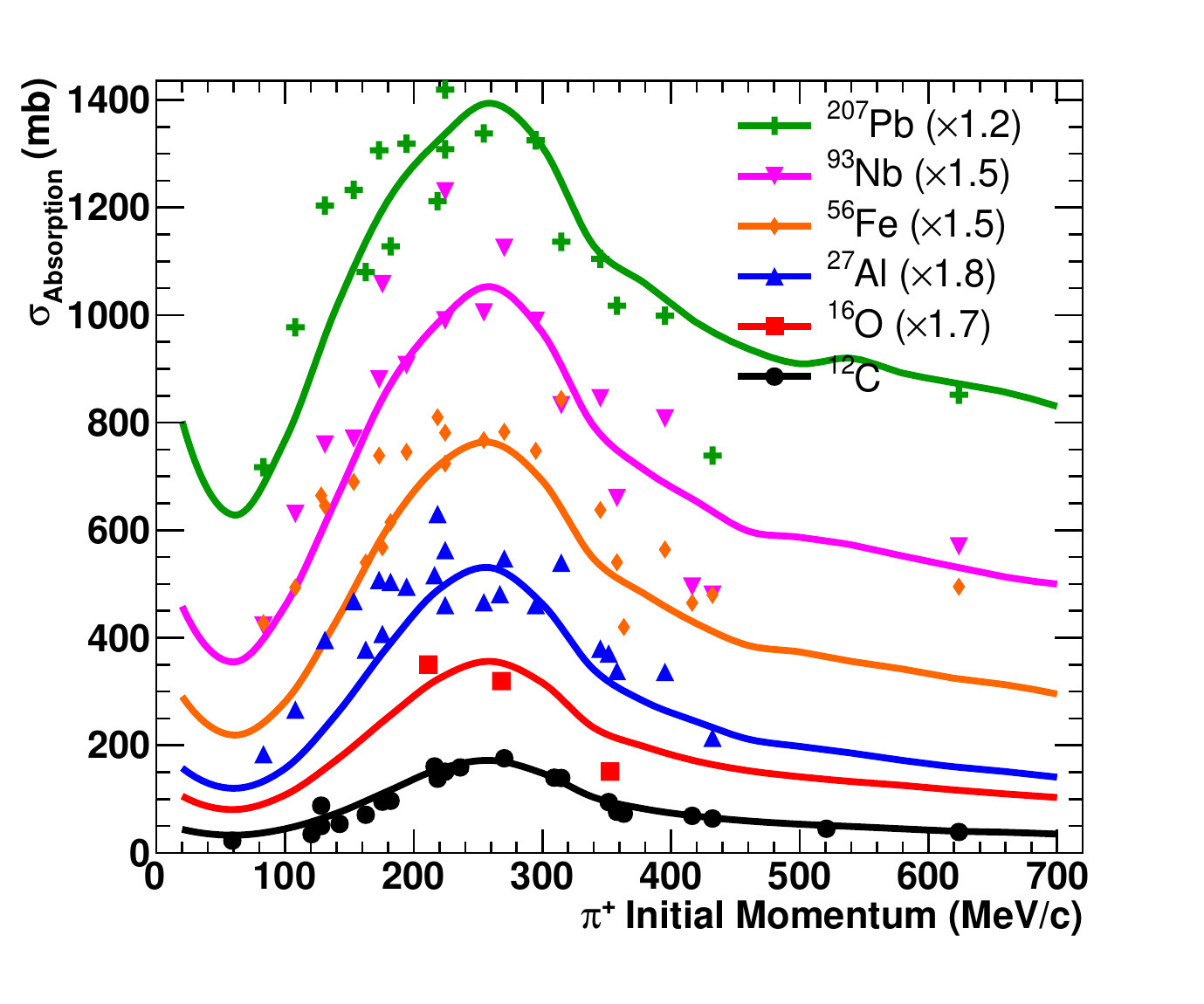}} \\
  \end{array}$  
  
  \caption
{Reactive (left) and absorption (right) cross sections for various target nuclei (tuned FSI).}
  \label{fig:piscat_ascale}
\end{figure}

\begin{figure}
     
  $\begin{array}{cc}
    \resizebox{3in}{!}{\includegraphics{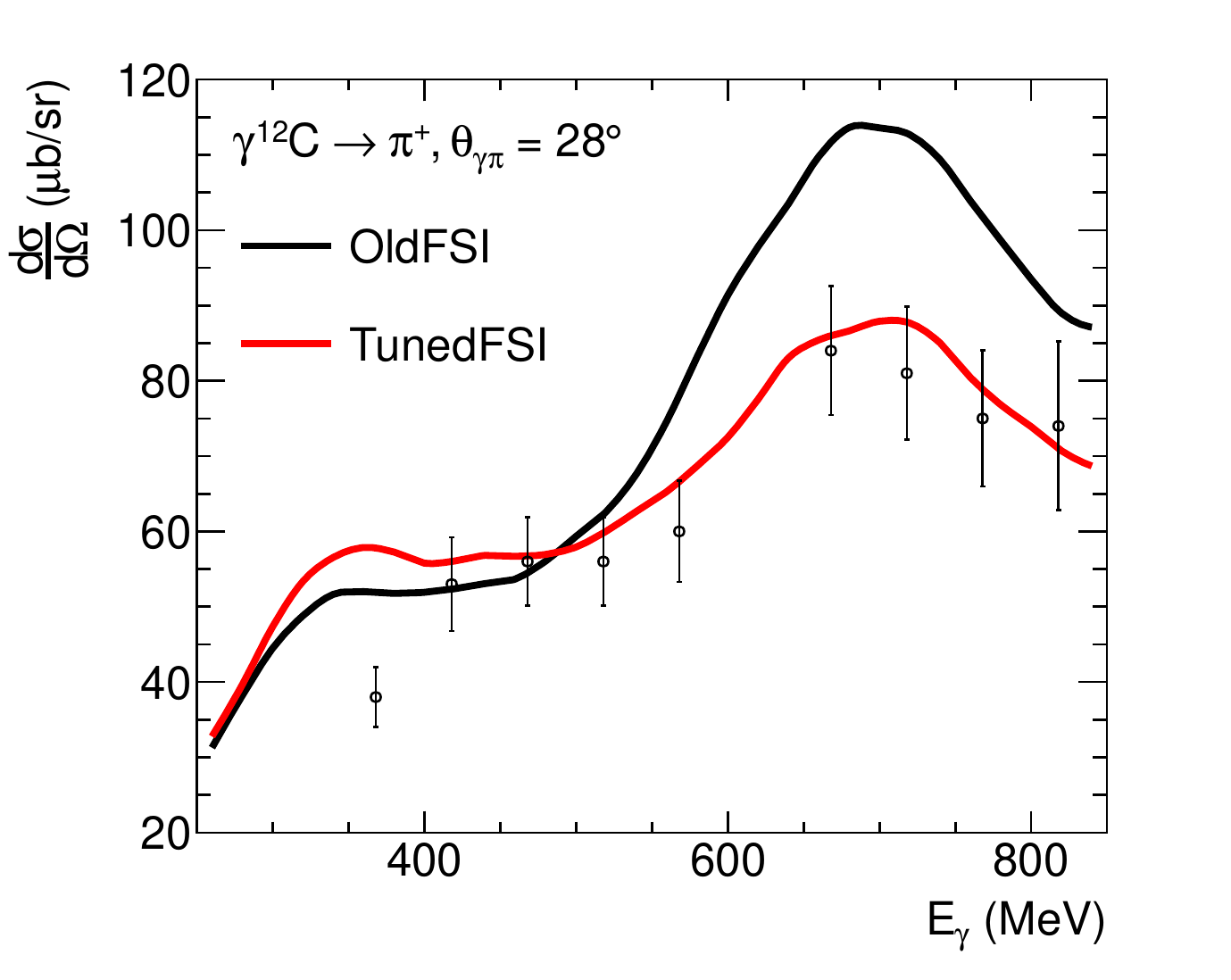}} &
    \resizebox{3in}{!}{\includegraphics{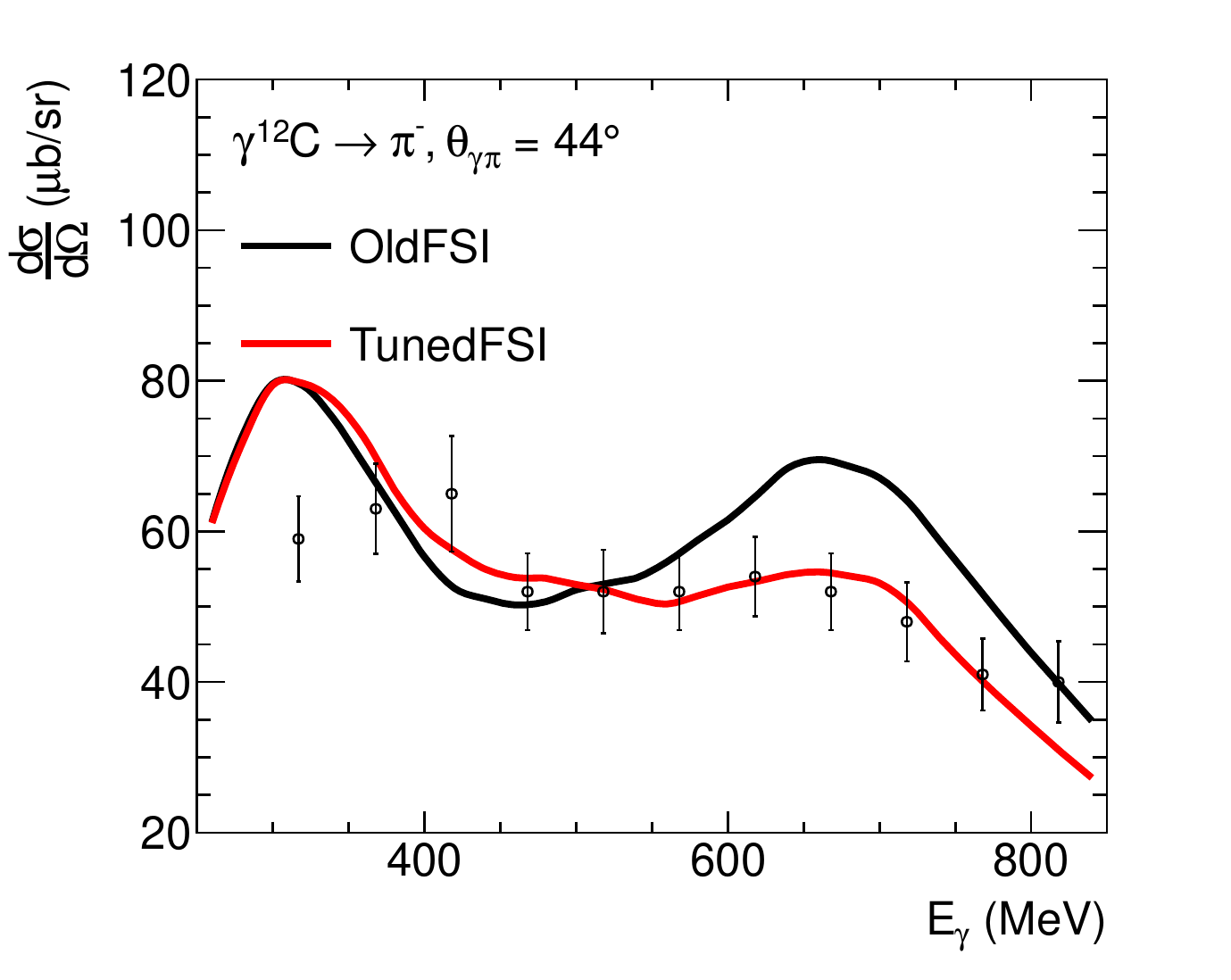}} \\
  \end{array}$  
  
  \caption
{Differential cross sections for $\pi$ photoproduction off carbon, as a function of photon beam energy. 
Data points from~\cite{baba:piphoto}.}
  \label{fig:photoprod_baba_dxsec}
\end{figure}

The assumption of an iso-scalar nucleus breaks down when considering heavier nuclei.
Hence, in the new model, the target nucleon is selected depending on the actual $p$/$n$ 
ratio of the nucleus, and the free $\pi^{\pm}p$ cross sections are used directly as
$\sigma_{QE}=\sigma_{QE}^{free}$, $\sigma_{had}=\sigma_{inel}^{free}$ and (a new)
$\sigma_{SCX}=\sigma_{SCX}^{free}$, as calculated by the SAID 
partial wave analysis (PWA) fit~\cite{pwa:pin} shown in Fig.~\ref{fig:freepxsec}. This also takes 
into account the different energy dependencies of each interaction 
mechanism depending on the initial state. Finally, we scale both $\sigma_{QE}$ and $\sigma_{SCX}$ 
by a constant factor of 1.8, as motivated by $\pi^{12}$C QE scattering~\cite{fujii:piscat} 
and SCX~\cite{jones:piscat} data, similar to the scaling factor measured in 
proton-nucleus scattering~\cite{bellettini:prot}.

The scattering kinematics at HE was originally determined from a 
simple elastic model: $\sin^2\theta_{CM} = -\frac{q_0^2\log{k}}{4|\mathbf{p_{\pi,CM}}|^2}$, 
where $q_0 = 200$~MeV/c and $k$ is a random number. This model lacks backward 
scattering, and a deficit in absorption and SCX at HE compared to data in 
Fig.~\ref{fig:piscat_c} also hints at this\footnote{More backward scattering would tend to 
reduce the $\pi$ energy into the $\Delta$ resonance region.}. Hence, we implement the 
phase shifts from SAID~\cite{pwa:pin} to calculate the differential cross section 
for HE $\pi{N}$ scattering in the new model.

The modifications to the HE region 
result in better agreement with data in all channels as shown in Fig.~\ref{fig:piscat_c}.
Blending of the LE and HE model was implemented to improve continuity at the boundary.
Pion scattering with heavier targets was also checked and we observe accurate 
reproduction of the $A$ dependence in each interaction channel as shown in 
Fig.~\ref{fig:piscat_ascale}, simply by varying the nuclear size and density 
distribution in the simulation. This provides confidence in the FSI model when 
simulating neutrino interactions on heavy nuclei, as in the T2K near detector (ND280)
or the surrounding rock of SK. We expect 95\% of $\pi$ produced with the T2K neutrino 
beam to have momenta $<2$~GeV/c.

A $\pi$-photoproduction simulation was developed, similar to that used in~\cite{arends:piphoto},
where the final state kinematics of the primary vertex are determined from a PWA fit of free 
nucleon photoproduction data~\cite{pwa:photoprod}. The results are compared to total, 
differential and double differential cross section data on carbon~\cite{baba:piphoto, arends:piphoto}.
For $E_{\gamma} > 500$~MeV, the addition of backward scattering to the FSI model, also resulting 
in more absorption, decreases the calculated cross section for forward going $\pi$ bringing 
the simulation into agreement with the data in Fig.~\ref{fig:photoprod_baba_dxsec}.

\begin{figure}
  
  $\begin{array}{cc}
    \resizebox{3in}{!}{\includegraphics{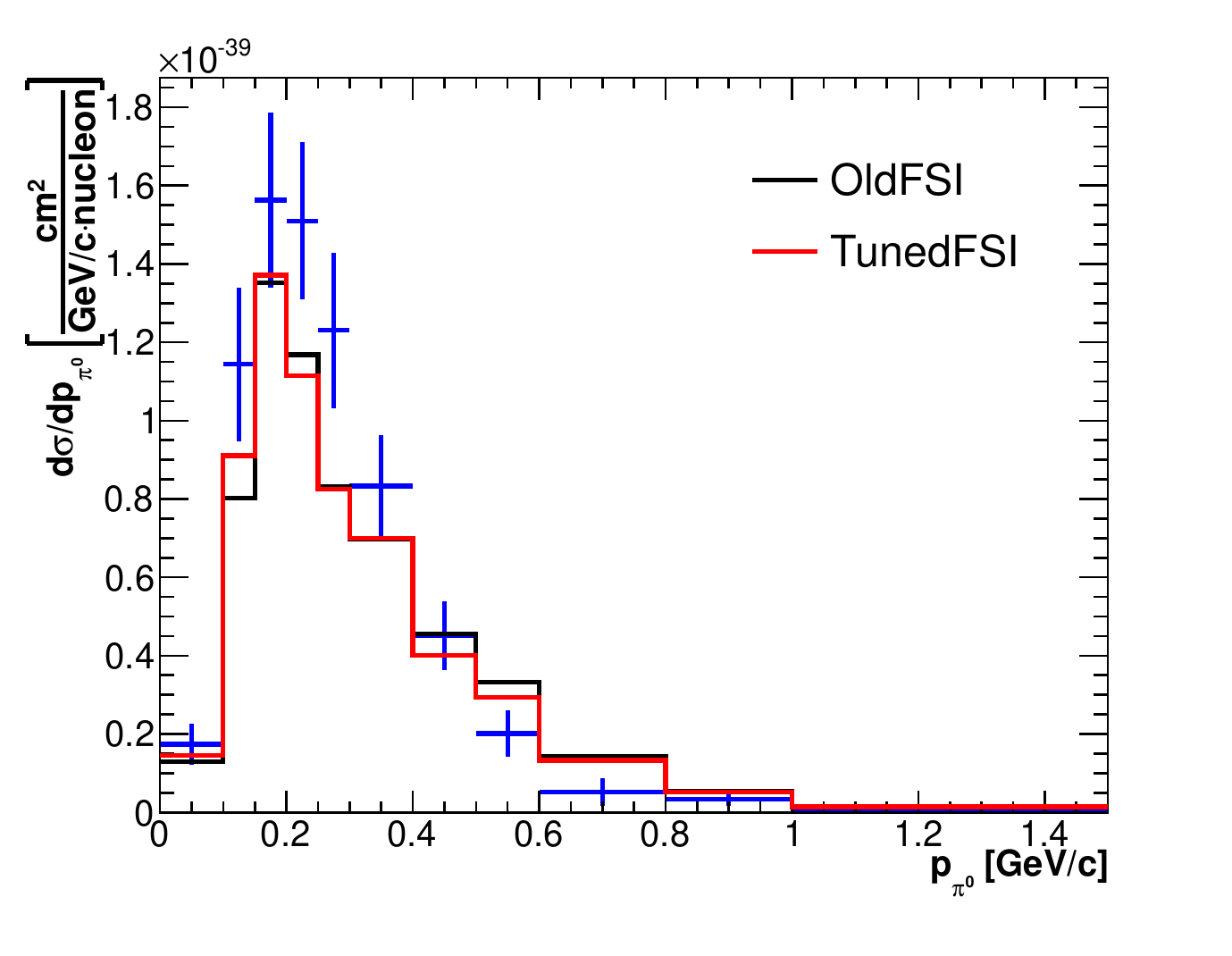}} &
    \resizebox{2.8in}{!}{\includegraphics{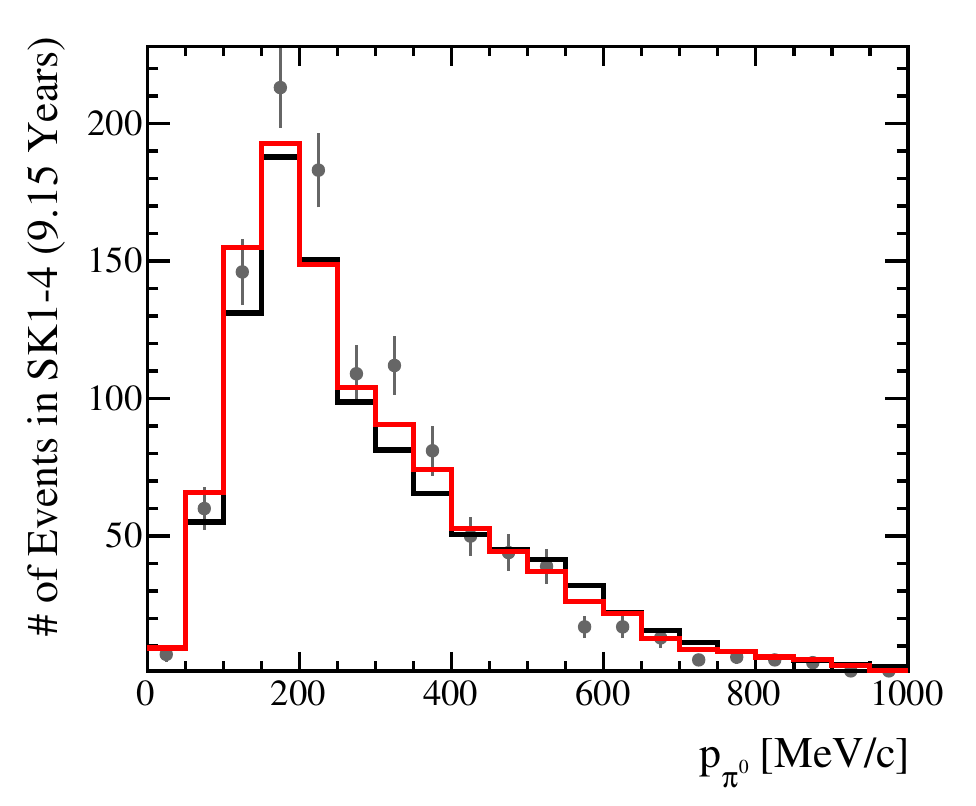}} \\
  \end{array}$  
  
  \caption
{NEUT predictions for the MiniBooNE~\cite{mb-nc1pi0} (left) and SK (right, 
statistical error only) NC, combined-sign, $\pi^0$ measurements.}
  \label{fig:neut_pi0}
\end{figure}

Finally, the full NEUT neutrino simulation was used with the MiniBooNE 
(MB) flux on CH$_2$ and compared to all available MB cross section data~\cite[e.g.][]{mb-nc1pi0}. 
Atmospheric neutrino interactions on H$_2$O at SK were also simulated, including the 
detector response and reconstruction~\cite{bib:t2k_experiment}. In Fig.~\ref{fig:neut_pi0}, 
we select NC events with a single outgoing $\pi^0$ for the MB comparison, and for SK, events with 2 
electron-like rings with reconstructed invariant mass between 85~and~185~MeV/c and no 
decay electron. There is slightly better agreement in the shape and absolute normalization 
after the FSI modifications.

\section{FSI Cascade Reweighting and Uncertainties}
\label{ch:fsirw}

Proper treatment of FSI systematics in neutrino experiments has 
been a longstanding problem. Variations in the final state would 
require regeneration of detector MC and reconstruction, a very CPU 
intensive task. To more easily predict changes in detector observables 
resulting from variations in microscopic MFPs, a reweighting scheme was developed
which preserves the details of the microscopic cascade and correlations between 
the different intranuclear mechanisms.

For each MC event we store the following 
truth information: $\pi$ starting and exit position,
position of each FSI vertex, mechanism at each vertex, $\pi$ type 
and momentum between each vertex and nucleus type. With this information, 
we can rerun the same cascade using the exact trajectory to calculate the 
probability $P_{evt}$ for the event. The survival probability 
in a given step of the cascade is: 
$P_{surv}(\mathbf{r},h,p) = 1 - \sum_iP_i(\mathbf{r},h,p)$,
where $\mathbf{r}$ is the current position within the nucleus, 
$h$ is the $\pi$ type, $p$ is the momentum and $i$ 
denotes the various interaction mechanisms. Then the probability 
for this given trajectory is: $P_{traj} = \prod_{all~steps}P_{step}$,
where $P_{step} = P_{surv}$ or $P_i$ depending on what occurred 
in the given step. The probability for an event with multiple 
$\pi$ is then: $P_{evt} = \prod_{all~trajs}P_{traj}$.

We define a set of energy and position independent scaling 
parameters, $f_i$, which scale each interaction probability, $P_i$. 
The modified probability of interaction for a given 
step is then $P_i' = f_iP_i$ which is used to recalculate 
$P_{evt}$. Finally, the weight for an event given a set of modified $f_i$
is: $w_{evt}(f_i) = \frac{P_{evt}'(f_i)}{P_{evt}}$. This weight can
be used when generating detector observable distributions, 
to observe the effect of varying the FSI model.

\begin{figure}
  
  $\begin{array}{cc}
    \resizebox{3in}{!}{\includegraphics{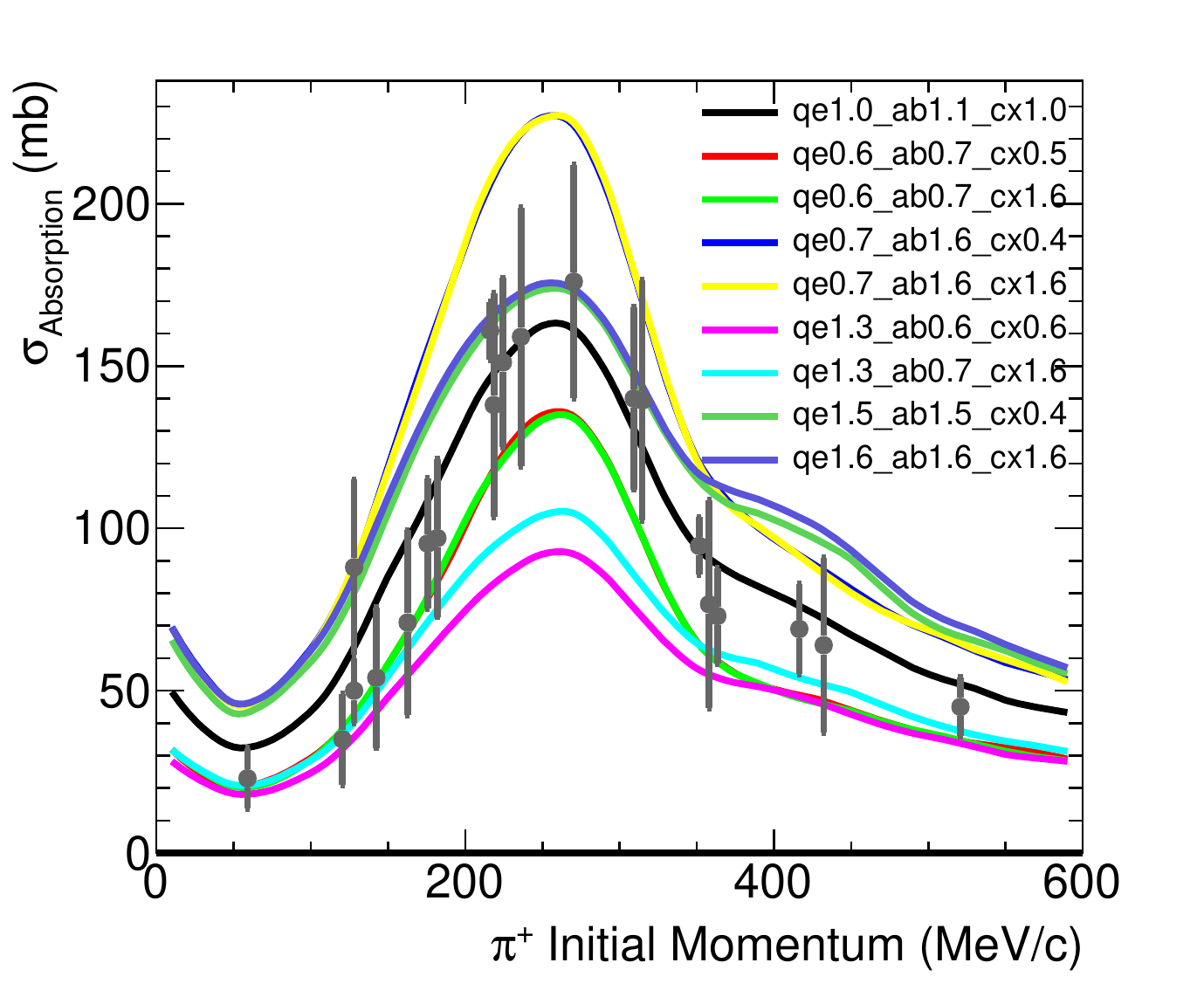}} &
    \resizebox{3in}{!}{\includegraphics{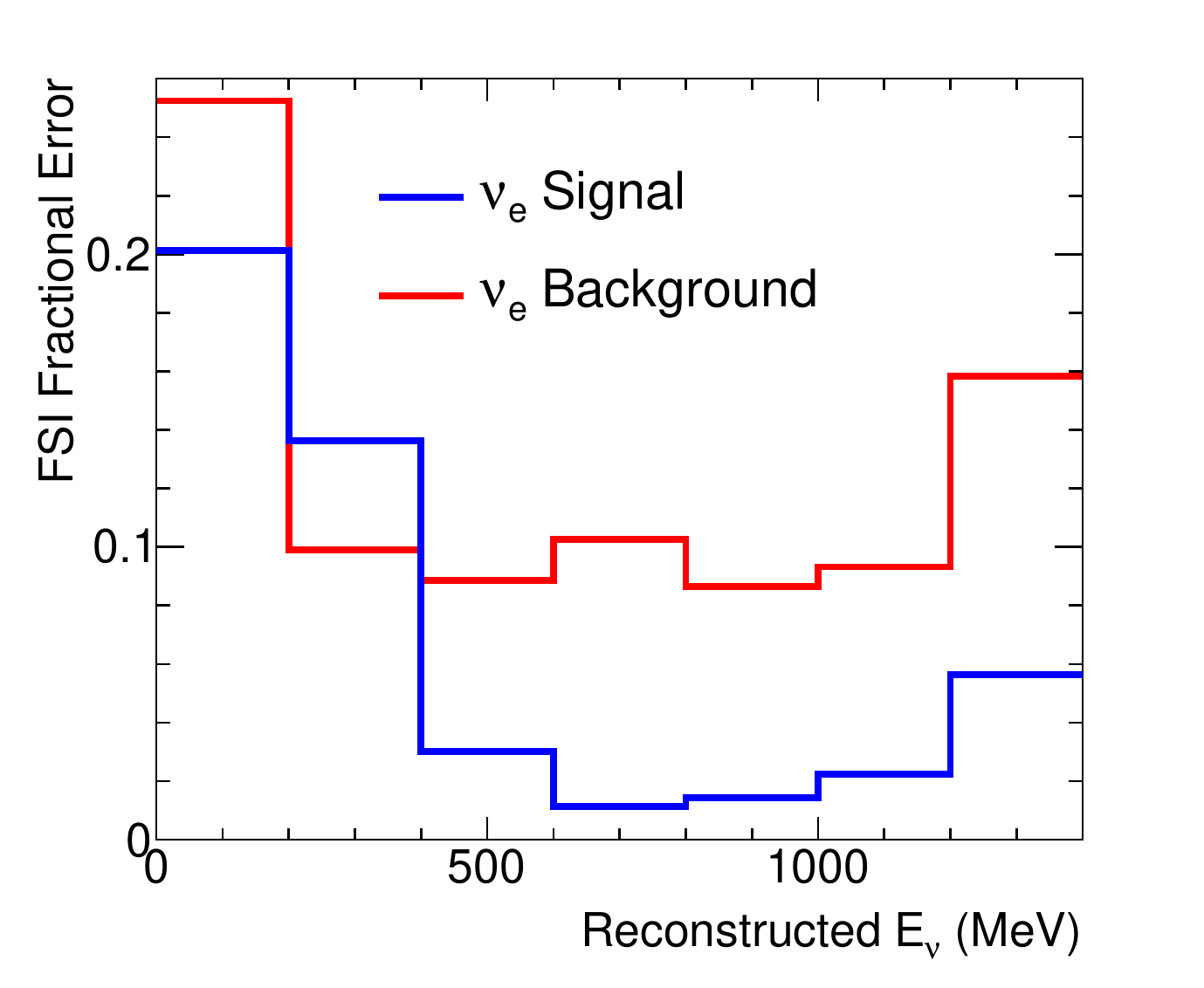}} \\
  \end{array}$  
  
  \caption
{Left: Tuned (black) and LE parameter variation $1\sigma$ curves from simultaneous fit to $\pi$ scattering data.
Right: Maximum deviation of the LE and HE FSI variations for the T2K-SK $\nu_{e}$ appearance 
(CCQE-like) sample, where ``$\nu_{e}$ Signal'' comes from CC interactions of $\nu_{e}$ 
oscillated from $\nu_{\mu}$~\cite{bib:t2k_nueapp1st}.}
  \label{fig:1sigpiscat}
\end{figure}

Three LE scaling parameters were simultaneously fitted to $\pi ^{12}$C data and 
the resulting eight 1$\sigma$ parameter sets (one from each octant of the 3-parameter 
space) are shown in Fig.~\ref{fig:1sigpiscat} (left). This set of parameters conservatively
spans the error of the data and is propagated through the T2K-SK and ND280 MC. The scaling 
parameters for HE were also varied for each LE set to maximize and minimize the event
multiplicity. Furthermore, we vary the secondary interaction cross sections in the SK 
detector simulation in a correlated manner with the intranuclear variations. 
For the T2K $\nu_{e}$ appearance analysis~\cite{bib:t2k_nueapp1st}, we simply use the maximum 
deviation across all FSI variations as the $1\sigma$ systematic error 
due to FSI uncertainties, as shown in Fig.~\ref{fig:1sigpiscat} (right). 
The increase in error towards low neutrino energy is due to non-CCQE events 
with all pions absorbed. For the ND280 CC inclusive measurement in~\cite{bib:t2k_nueapp1st}, 
the effect is less significant ($<1\%$) since it is insensitive to FSI. However, future
exclusive selections will benefit from this reweighting scheme. Also, future oscillation 
analyses will be able to properly handle correlations introduced by FSI uncertainties
between near and far detectors.

\section{Summary and Acknowledgments}

Modifications to the NEUT FSI model, necessary for extracting a constraint from $\pi$ scattering data, 
were described. The model shows good agreement compared to external pion scattering, photoproduction 
and neutrino data. A reweighting scheme was developed for evaluating systematic errors due to FSI 
uncertainties, with an example application to the T2K experiment.

I am grateful for the tutelage of my colleagues at Super-Kamiokande, especially 
Yoshinari Hayato. I acknowledge the support of the NSERC MSFSS, CGS and Vanier programs.






\bibliographystyle{aipproc}   

\bibliography{neut_pion_fsi}

\IfFileExists{\jobname.bbl}{}
 {\typeout{}
  \typeout{******************************************}
  \typeout{** Please run "bibtex \jobname" to obtain}
  \typeout{** the bibliography and then re-run LaTeX}
  \typeout{** twice to fix the references!}
  \typeout{******************************************}
  \typeout{}
 }

\end{document}